# Probing picosecond depairing currents in type-II superconductors

E. Wang[1†], M. Chavez-Cervantes[1], J. Satapathy[1], T. Matsuyama[1], G. Meier[1], X. Zhang[2], L. You[2], F. Marijanovic[3], J.B. Curtis[3], E. Demler[3], A. Cavalleri[1,4†]

1 Max Planck Institute for the Structure and Dynamics of Matter, Hamburg, Germany

2 Shanghai Key Laboratory of Superconductor Integrated Circuit Technology, Shanghai Institute of Microsystem and Information Technology, Chinese Academy of Sciences, 200050 Shanghai, China

3 Institute for Theoretical Physics, ETH Zurich, 8093 Zurich, Switzerland

4 Clarendon Laboratory, University of Oxford, Parks Road, Oxford OX1 3PU, United Kingdom

† Corresponding authors: eryin.wang@mpsd.mpg.de; andrea.cavalleri@mpsd.mpg.de

**Accessing the intrinsic critical current density ($J_c^*$) in type II superconductors has significant fundamental and technological potential, both as a probe of the microscopic superconducting properties and as a means to increase current limits in high magnetic field devices and in electrical power systems. Yet, the experimental critical current density in type II superconductors ($J_c$), when measured with DC currents, is generally lower than the intrinsic limit, mostly due to vortex motion and self-heating. Here, we show that ultrafast picosecond electrical pulses, which interact with the material on timescales over which vortices are inertially immobile, carry supercurrents up to the intrinsic depairing limit $J_c^* \gg J_c$. We probe picosecond critical currents in NbN and YBa$_2$Cu$_3$O$_7$ (YBCO), as representative *s*-wave and *d*-wave superconductors, respectively. In NbN, we find a sharp onset of the picosecond depairing at a current density as large as $J_c^* = 2.2 \times J_c$, a limit that is well described by microscopic dynamics based on BCS theory. In contrast, YBCO exhibits a gradual suppression of superconductivity as a function of the picosecond current, reflecting its *d*-wave symmetry. These results offer a powerful new probe of superconductors beyond the reach of conventional transport measurements. The ability to reach the depairing current may also lead to robust new platforms for superconducting electronics.**

The amplitudes of critical quantities in superconductors, namely the critical temperature ($T_c$), critical magnetic field ($H_c$), and critical current density ($J_c$), are essential observables that yield insight into the physics of these materials, as well as important figures of merit for their application. In this context, type-II superconductors have attracted greater interest due to their high $T_c$ values compared to type-I superconductors. In type-II systems, the magnetic penetration depth ($\lambda$) is larger than the superconducting coherence length ($\xi$), which leads to the formation of magnetic vortices. Consequently, some thermodynamic critical quantities are split into lower and upper critical values. For example, the critical magnetic field divides into the lower and upper critical fields, denoted as $H_{c1}$ and $H_{c2}$ (Fig.1a), respectively.

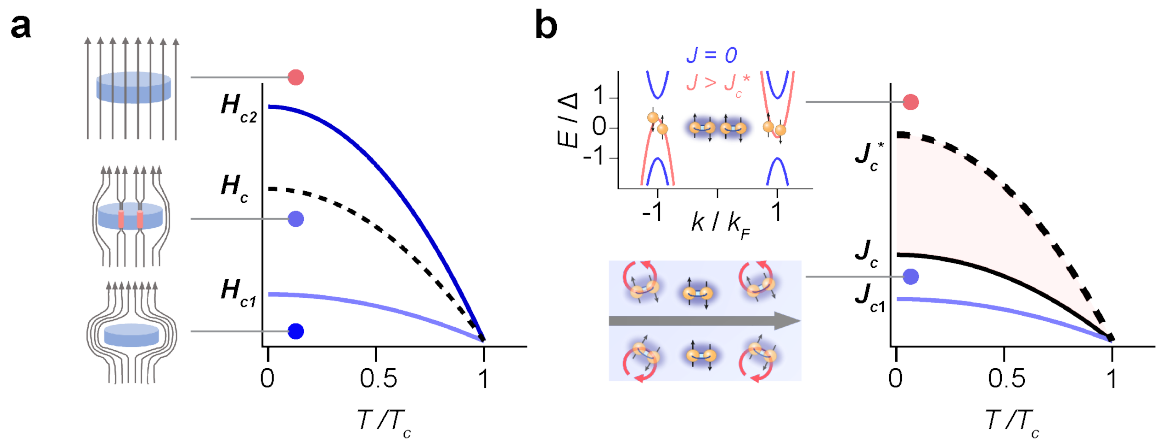

**Figure 1| Critical magnetic fields and current densities in type-II superconductors.** (a) Critical magnetic fields in type-II superconductors. $H_{c1}$ and $H_{c2}$ are the lower and upper critical magnetic fields. Above $H_{c1}$, magnetic vortices penetrate the superconductor. Above $H_{c2}$, superconductivity is totally suppressed. $H_c$ is the thermodynamic critical magnetic field at which the magnetic field expulsion energy cost equals the condensation energy gain. Due to the formation of vortices above $H_{c1}$, $H_c$ is inaccessible experimentally. (b) Critical current densities in type-II superconductors. $J_{c1}$ corresponds to the cases where the self-magnetic field of current reaches $H_{c1}$ and magnetic vortices penetrate sample (left-bottom, vortices indicated as oversized red circles for clarity). $J_c^*$ is the thermodynamic depairing current density at which the quasiparticle energy shift exceeds the gap size $\Delta$, and Cooper pairs start depairing (left-top, quasiparticle energy dispersion with $J = 0$ and $J > J_c^*$ shown in blue and red colors). However, the depinning of vortices and subsequent self-heating at current densities above the conventional critical current density $J_c$ inhibit the observation of $J_c^*$ and this conventionally inaccessible region is indicated by the shaded red color.

When the applied magnetic field $H$ exceeds $H_{c1}$, vortices – each with characteristic size on the order of the coherence length $\xi$ – begin to penetrate into the superconductor (Fig. 1a). As the applied magnetic field $H$ increases further, more vortices enter the material while the sample remains superconducting. Superconductivity persists until the field surpasses $H_{c2}$, at which point vortices occupy the entire sample and superconductivity is fully suppressed [1]. The thermodynamic critical field $H_c$ is defined as the field at which the magnetic field expulsion energy $(H_c^2/4\pi) \times V$ equals the free energy difference between normal and superconducting states $(F_N - F_S)$, and is located between $H_{c1}$ and $H_{c2}$. Due to the formation of vortices above $H_{c1}$, magnetic field is not fully expelled around the sample and thus $H_c$ is experimentally not accessible.

Similarly, in superconducting electronics the maximum current a superconductor can sustain is also a critical parameter. This limit is defined by the *conventional* critical current density $J_c$, set by vortex depinning through the Lorenz force (Fig. 1b), which leads to resistive heating and a subsequent transition to the normal state [2-7]. However, $J_c$ is not an intrinsic material property associated with the microscopic superconducting parameters, but rather strongly depends on the defect density within the material, which determine the vortex pinning potential.

A thermodynamic depairing current density denoted as $J_c^*$, can be defined when the quasiparticle energy shift $\hbar k_F \cdot v_s$ equals the superconducting gap $\Delta$ (here $k_F$ is the Fermi momentum and $v_s$ is the velocity of Cooper pairs, see Fig. 1b) [1]. Above $J_c^*$, superconductivity is no longer the energetically favored state and Cooper pairs dissociate into normal carriers. $J_c^*$ is therefore an intrinsic property of the superconducting material and sets the fundamental upper limit for the supercurrent.

Although the depairing mechanism has been theoretically predicted and experimentally suggested [8-15], direct transport measurements of the depairing current density $J_c^*$ remain challenging. When a DC current density exceeds $J_c$, vortex penetration and subsequent self-heating set in over timescales on the order of nanoseconds [16-18], driving the sample into the normal sate before $J_c^*$ can be observed.

Here, we use a picosecond ultrafast electrical transport platform to investigate the dynamics of type-II superconductors subjected to strong, ultrashort current pulses, focusing on both *s*-wave and *d*-wave superconductors. The core idea is that within picosecond timescales, the penetration of vortices is inertially frozen near the edges, as their speed is intrinsically limited to 10s of nm/ps (10s of km/sec) [7, 18-22], which leads to negligible energy dissipation and leaves the bulk of the sample unaffected.

## DC and picosecond electrical transport measurements

For this study, we choose a representative *s*-wave superconductor, NbN, and a representative *d*-wave superconductor $YBa_2Cu_3O_7$ (YBCO). Detailed device fabrication procedures are provided in the Supplementary Information.

The DC transport properties of both NbN and YBCO samples were first characterized, as shown in Fig. 2. Conventional critical current densities of approximately 100 GA/m$^2$ and slightly below 50 GA/m$^2$ were observed for NbN at 7 K and YBCO at 50 K, respectively.

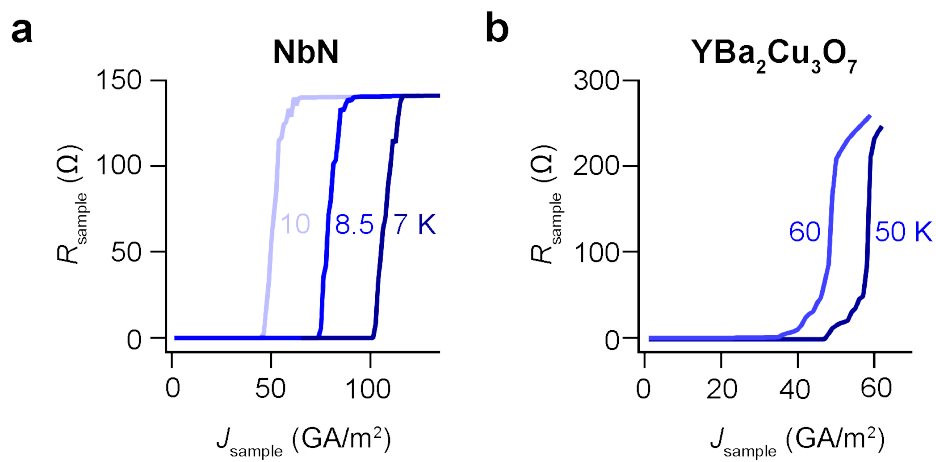

**Figure 2| Conventional critical current densities $J_c$ in s-wave and d-wave type-II superconductors.** (a) Measurements of $J_c$ in the *s*-wave superconductor NbN at 7, 8.5 and 10 K, with $T_c$ = 14.5 K. (b) Measurements of $J_c$ in the *d*-wave superconductor YBCO at 50 and 60 K, with $T_c$ = 85 K.

Figure 3 displays the picosecond electrical transport measurements on the s-wave superconductor NbN; analogous measurements for the d-wave superconductor YBCO are provided in the Supplementary Information. The device architecture is shown in

Fig. 3a. A NbN thin film, approximately 20 nm thick, is connected to three pairs of photoconductive switches with a coplanar waveguide.

Picosecond current pulses (FHMW ~ 2 ps) are launched by illuminating the middle pair of voltage-biased photoconductive switches with 515-nm, femtosecond laser pulses [23-28]. The pulse is referenced using one of the unbiased switches on the left-hand side of the sample, without interaction with the superconductor. Reflected and transmitted pulses are then sampled at later time delays using the same unbiased switch on left hand side and the one on the right-hand side, respectively. Details of the calibration procedure are provided in the Supplementary Information. Figure 3b shows the normalized incoming ($E_{IN}$), reflected ($E_R$) and transmitted ($E_T$) electric field at $T$ = 7 K and 20 K, corresponding to temperatures below and above the superconducting critical temperature $T_c$ = 14.5 K. The electric field $E(t)$ is directly related to the current density $J(t)$ through the expression $E(t) \cdot w = S \cdot J(t) \cdot Z_0$, where $w$ is the gap between the ground plane and signal line, S is the sample cross section area, and $Z_0 \approx 50\ \Omega$ is the wave impedance of the coplanar waveguide.

At $T$ = 7 K, the reflected pulse exhibits a small inductive response proportional to $L_{kin} \cdot dI_{T,ps}(t)/dt$, where $L_{kin}$ is the kinetic inductance of the sample and $I_{T,ps}(t)$ is the transmitted current pulse. The majority of the incoming pulse is transmitted, with the transmitted peak reaching approximately 90% of the incident peak. In contrast, at 20 K, the pulses are partially reflected and transmitted due to the resistive behavior of the sample in the normal state.

Figure 3c displays the response of the NbN sample at $T$ = 7 K under both weak (~ 0.2 × $J_c$) and strong (~ 6.5 × $J_c$) current pulse drive, where $J_c$ is the DC critical current density measured at 7 K. These values correspond to the peak current amplitudes of the transmitted current pulses for comparative analysis.

Under weak current pulse drive, the sample response is consistent with the previously described superconducting behavior. However, under strong current pulse drive, the response closely resembles the one observed for small amplitude pulses in the normal state at 20 K. This observation suggests that strong picosecond current pulses can transiently suppress superconductivity, likely through the instantaneous depairing of Cooper pairs into normal carriers.

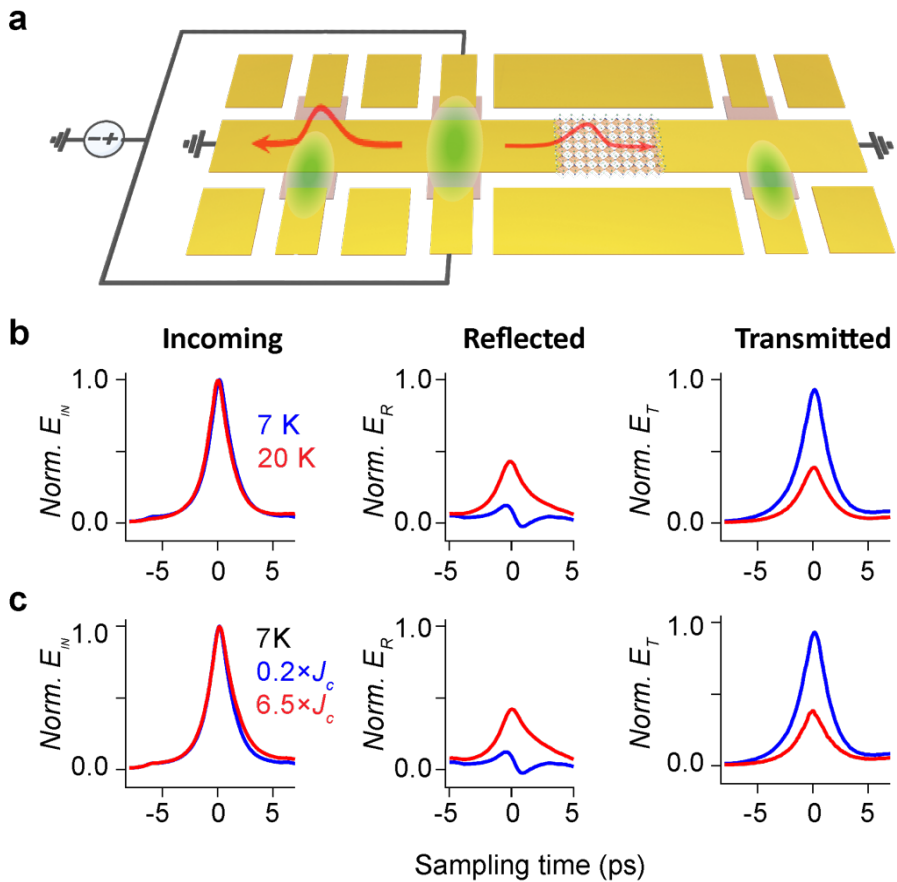

**Figure 3| Picosecond ultrafast electrical transport measurements on NbN.** (a) Illustration of device architecture. The picosecond current pulses are launched from the middle pair of voltage-biased photoconductive switches (grey patches) with illuminating 515-nm laser beams. The incoming/reflected and transmitted pulses are sampled with one of the left and right pairs of unbiased switches, respectively. (b) Measurements at 7 K and 20 K, below and above $T_c$ = 14.5 K, respectively. At 7 K, the reflected pulses show an inductive feature and ~90% of the incoming pulses transmits through. At 20 K, the incoming pulses partially reflect and partially transmit through, due to sample's resistive response. (c) Measurements at 7 K with different peak current densities of incoming pulses. Here the labels indicate the peak current densities of the transmitted pulses for direct comparison. At 6.5×$J_c$, the response closely resembles the sample's response at 20 K, indicating strong suppression of superconductivity. All curves shown here are normalized by the peak electrical field of incoming pulses.

## Observation of intrinsic depairing currents

The transition dynamics of type-II superconductors under strong picosecond current drive are shown in Figures 4a and 4b for the *s*-wave superconductor NbN. The electric field transmittance (*Norm. $E_{T,peak}$*) — peak amplitude of the transmitted electric field normalized by that of the incoming pulse — is plotted versus the peak current density flowing in the sample ($J_{T,peak}$) to track the superconducting state.

As shown in Figure 4a, at $T$ = 7 K, the normalized transmitted peak field $E_{T,peak}$ remains constant above the DC critical current density $J_c$, and then exhibits a sharp drop at around 2.2 × $J_c$, followed by a gradual transition towards the normal-state response, measured at 20 K. The slight leftward tilt observed during the sharp drop can be attributed to the reduction in the transmitted peak current amplitude, which will be discussed further below. Figure 4b plots the transmitted peak current density as a function of the incoming peak current density. At both weak and strong driving current pulses, the response at 7 K displays a linear relationship. The slope in the high-current regime closely matches that of the normal state at 20 K, where current is led by normal charge carriers.

These two linear regions correspond to regimes in which the transmitted current is predominantly carried by Cooper pairs and by normal carriers, respectively. Within the intermediate regime, increasing the incident current results in a reduction of the transmitted peak current, corresponding to the slight leftward tilt shown in Fig. 4a noted above. This pronounced decrease in supercurrent is attributed to the rapid depairing of Cooper pairs [1]. A local maximum in transmitted peak current density — highlighted in the inset of Figure 4b — marks the depairing current density $J_c^*$ introduced above.

Measurements on the *d*-wave superconductor YBCO are shown in Figures 4c and 4d. In contrast to the results on NbN, the normalized transmitted peak electric field $E_{T,peak}$ in YBCO exhibits a gradual and continuous decrease towards the normal state response at 110 K — both before and after $J_{T,peak}$ reaches the DC critical current density $J_c$. Similarly, as shown in Figure 4d, the slope of the transmitted peak current density versus the incoming peak current density continuously converges to the value characteristic of the normal state. Notably, unlike in NbN, no local maximum in the transmitted peak current density was observed. These results suggest that in *d*-wave type-II superconductors, in which the superconducting gap evolves from zero at the nodal point to its maximal value at the anti-nodal point, there is no well-defined intrinsic current density threshold above which Cooper pairs begin to depair. Instead, partial depairing occurs continuously as current is applied.

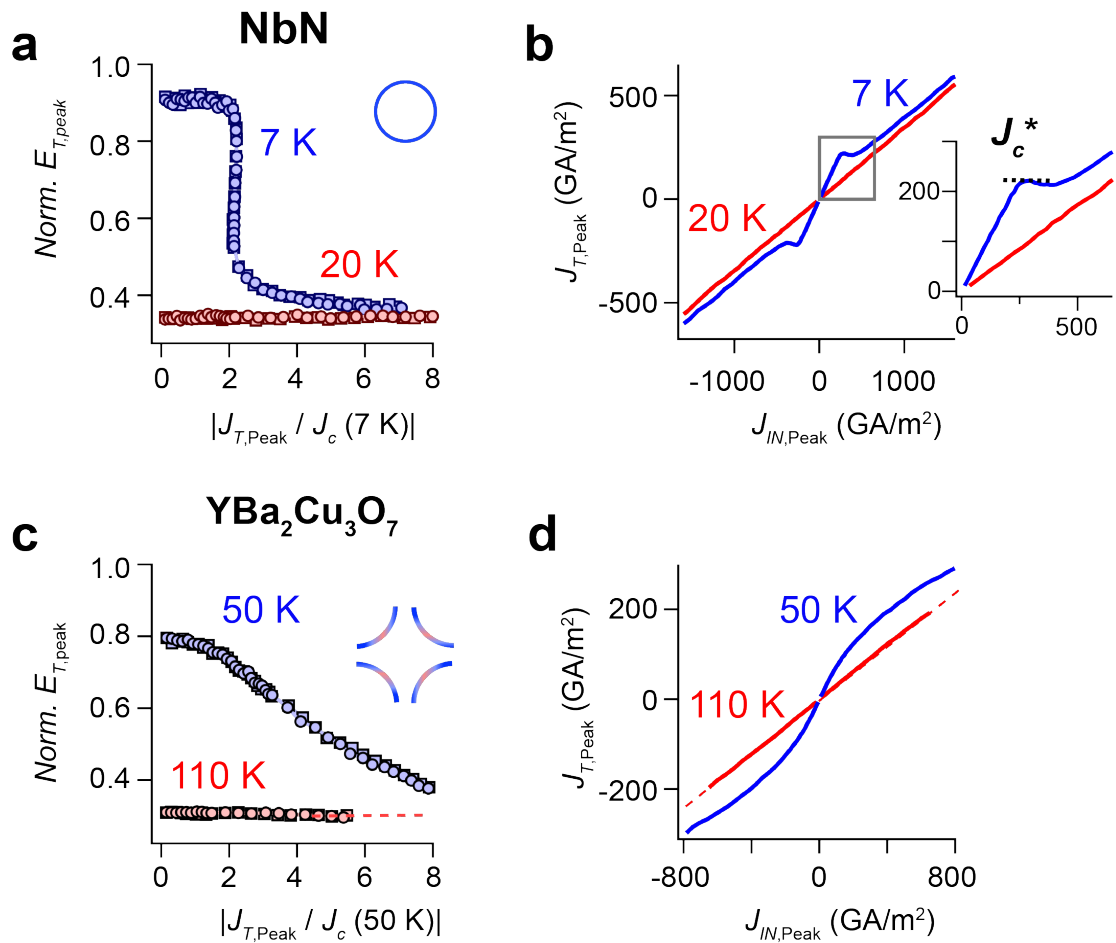

**Figure 4| Intrinsic depairing current density $J_c^*$ in *s*-wave and *d*-wave type-II superconductors.** (a-b) Picosecond electrical transport measurements on the s-wave superconductor NbN. (a) Normalized transmitted electric peak field (by the peak electric field of the incoming pulses) versus the transmitted peak current densities at 7 K and 20 K, respectively. A sharp drop is observed at 7 K in contrast to the flat feature of the normal state at 20 K, indicating a sudden suppression of superconductivity. The inset blue circle indicates the isotropic s-wave gap. ● and ■ symbols represent the cases with positive and negative picosecond pulses. (b) Transmitted peak current density versus incoming peak current density. In contrast to the linear relationship at 20 K, the transmitted peak current density starts dropping at ~ 220 GA/m² and then starts increasing again with the similar slope of the normal state, indicating a sudden replacement of Cooper pairs with normal carriers. (c-d) Picosecond electrical transport measurements on the d-wave superconductor YBCO. (c) Same as (a). In contrast with NbN, the normalized transmitted peak field shows an immediate and continuous drop with increasing transmitted peak current density at $T$ = 50 K below $T_c$ = 85 K, indicating continuous depairing of Cooper pairs. (d) Same as (b). The slope of the measured curve at 50 K shows a slow reduction, consistent with the observation in (c).

## Discussion

We first rule out the possibility that the observed effects arise from depinned magnetic vortices. Firstly, all experiments were conducted under zero-field cooling conditions, with the ambient magnetic field reduced to below 1 μT using dual magnetic shielding. Such conditions ensure that no pre-existing magnetic vortices are present within the sample area (width ≈ 10 um × length ≈ 30 um). Secondly, even when neglecting the finite time required for vortex entry and depinning under picosecond current drive, the maximum displacement of a vortex would be less than a few tens of nm, given maximum vortex velocities on the order of a few tens of kilometers per second [7, 18-22]. Full vortex penetration is therefore implausible. This explains why vortex-induced resistive responses in superconductors typically manifest only on nanosecond timescales [16-18]. Thirdly, the distinct responses observed in the *s*-wave superconductor NbN and the *d*-wave superconductor YBCO suggest that the observed picosecond suppression of superconductivity is closely related to the microscopic gap symmetry.

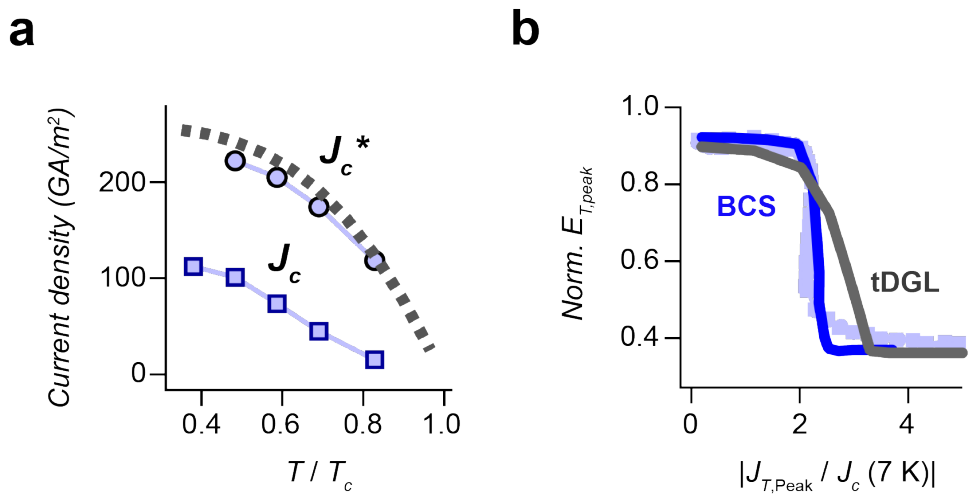

**Figure5| Calculations of depairing current density $J_c^*$ and simulations of depairing process in the *s*-wave superconductor NbN.** (a) The measured conventional critical current density $J_c$ and intrinsic depairing current density $J_c^*$ versus temperature are shown as ■ and ● symbols, respectively. The calculated $J_c^*$ versus temperature curve is shown as a grey dashed line. (b) Simulations of the depairing process in NbN at 7 K based on BCS theory in dirty limit and tDGL theory, which are show in blue and grey curves. The measurement data at 7 K is shown in light blue.

Taken together, these three considerations support the conclusion that the observed depairing current density $J_c^*$ in the *s*-wave type-II superconductor NbN originates from the competition between quasiparticle energy shift and the superconducting gap. In superconductors with an anisotropic gap in momentum space, such as *d*-wave superconductors, $J_c^*$ is not a well-defined quantity and is therefore not observable in the ultrafast transport measurements on YBCO. We note that, the monocrystallinity of both NbN and YBCO thin films (see supplementary information) is important here. In polycrystalline superconductor thin films, the phase slip dynamics at domain boundary dominate the sample's nonlinear response [26] and impede the observation of intra-domain depairing processes.

Figure 5a compares the conventional critical current density $J_c$ and the intrinsic depairing current density $J_c^*$ as functions of temperature (raw data is provided in Supplementary Information). As the temperature increases, $J_c$ decreases significantly faster than $J_c^*$; at 0.8 × $T_c$, $J_c$ is nearly an order of magnitude smaller than $J_c^*$.

Since $J_c^*$ is fundamentally determined by the microscopic properties of the *s*-wave superconductor, it can be calculated directly from microscopic parameters. Assuming at $J_c^*$, the quasiparticle energy shift $\hbar \cdot k_F \cdot v_s$ equals the superconducting energy gap Δ, yielding:

$$v_s = \frac{\Delta}{\hbar \cdot k_F} = \frac{\hbar}{\pi m \xi}$$, where m is the effective mass of the electron and ξ is the superconducting coherence length.

The depairing current density is then given by

$$J_c^* = n_s v_s \cdot (2e) = \frac{2\hbar e}{\pi m} \frac{n_s}{\xi},$$

where $n_s$ is the superfluid density. By substituting the temperature-dependent functions $n_s(T)$ and ξ (*T*) (see supplementary information), the calculated temperature dependence of $J_c^*(T)$ shows excellent agreement with the experiments, with the experimentally measured values slightly smaller than the calculated ones. This is possibly because the depairing process partially starts before the quasiparticle energy shift $\hbar \cdot k_F \cdot v_s$ reaches the superconducting energy gap Δ due to thermal fluctuations. Details of the calculation are provided in the Supplementary Information.

We simulated the depairing process in the *s*-wave superconductor NbN, using both time-dependent Ginzburg-Landau (tDGL) theory and microscopic BCS theory in the dirty limit relying on the Usadel equation [29]. The microscopic theory directly takes parameters from experimental characterizations of the sample (details see supplementary information), apart from Dynes broadening, which is set to guarantee numerical convergence and to reproduce the experimental results. In general, the sharp transmission drop is attributed to the strong-nonlinearity of the superfluid density, which depends on the s-wave nature and strong disorder of the system. tDGL cannot capture the strong non-linearity necessary to model this behavior, since it relies on expanding the free energy in powers of the order parameter and is only valid near $T_c$ [30, 31]. This consistency further confirms that the observed sudden change in the sample's response originates from the rapid depairing of Cooper pairs in *s*-wave superconductors.

In summary, we have reported ultrafast electrical transport measurements in which vortex motion and subsequent self-heating are bypassed—thereby enabling the extraction of intrinsic microscopic properties of type-II superconductors. The findings reported here also offer a new approach to probing the gap symmetry in systems that cannot be directly accessed using spectroscopic techniques. From an applications perspective, our results further reveal how maximum supercurrents in a type-II superconductor can be reached on a picosecond timescale, providing a potential platform for generating ultrashort, strong magnetic field pulses.

# Supplementary information for
# Probing picosecond depairing currents in type-II superconductors

## S1. Sample characterizations

### S1.1 Characterization of the NbN thin film

NbN samples were deposited by dc magnetron sputtering on MgO substrates with a substrate temperature of 300 °C, with a thickness of ≈ 20 nm and a superconducting critical temperature of $T_c$ ≈ 14.5 K [1]. The thin films were characterized by XRD shown in Fig. S1. A clear separation between the Bragg peaks of NbN and MgO can be observed at $(\bar{6}00)$ and $(\bar{4}00)$. The other Bragg peaks cannot be clearly separated because of the strong intensity of the peaks from MgO, and merge as stripes stretched towards the origin as expected for such a configuration. This confirms the epitaxial growth of monocrystalline NbN thin films.

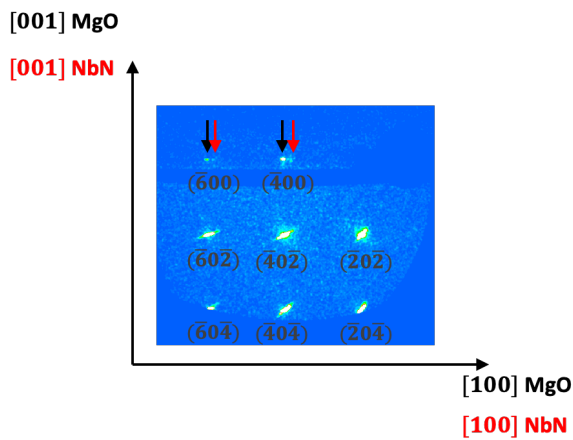

**Figure S1| XRD pattern of the NbN film grown on MgO.** NbN and MgO both have cubic lattice structure but with different lattice constants ($a_{NbN}$ ≈ 4.4 Å and $a_{MgO}$ ≈ 4.2 Å). A clear separation between the Bragg peaks of NbN and MgO can be observed at $(\bar{6}00)$ and $(\bar{4}00)$. The other Bragg peaks cannot be clearly separated because of the strong intensity of peaks from the MgO substrate. Instead, these Bragg peaks are stretched towards the origin as expected for such a configuration.

### S1.2 Characterization of the YBCO thin film

YBa$_2$Cu$_3$O$_7$ (YBCO) samples were deposited by dc magnetron sputtering on sapphire substrates with 10 nm CeO$_2$ buffer layer, with a thickness of ≈ 50 nm and a critical temperature of ≈ 85 K by Ceraco ceramic coating GmbH. 50 nm Au was deposited on top of YBCO in situ after sputtering to get good electrical contact. The bare YBCO samples were characterized by XRD shown in Fig. S2. This confirms the epitaxial growth of monocrystalline YBCO thin film.

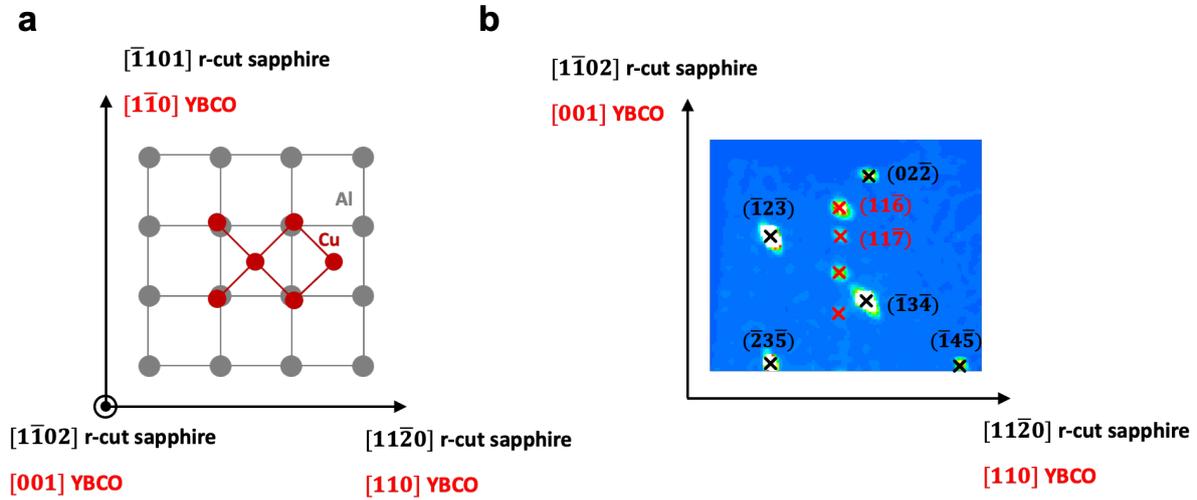

**Figure S2| XRD pattern of YBCO film grown on r-cut sapphire with CeO$_2$ buffer layer.** a, Illustration of the arrangement of Cu atoms from YBCO on Al atoms from Al$_2$O$_3$. b, Bragg peaks in $(1\bar{1}0)$ plane of YBCO. The peaks indicated in black correspond to the r-cut sapphire while the red peaks relate to YBCO. Because the thickness of the CeO$_2$ buffer layer is much thinner than YBCO, its pattern is not observable.

## S2. Device fabrication procedures

NbN thin films were patterned with laser lithography and reactive ion etching (RIE) using SF$_6$. Photoconductive switches and coplanar waveguides were fabricated on MgO substrates with pre-patterned NbN thin films using laser lithography and e-beam deposition. Specifically, 200 nm thick silicon was deposited to form photoconductive switches and 10 nm Ti / 260 nm Au was deposited to form coplanar waveguides. In order to get good electrical contact between the waveguide and the patterned NbN thin film, the contact area of the NbN thin film was cleaned with Argon plasma and was then capped with 50 nm Au in situ, before depositing Ti/Au to form a waveguide. With such procedures, a contact resistance of ≈ 3 Ω for each contact was obtained.

YBCO together with the top Au layer was patterned with laser lithography and wet etching methods. A solution of KI and I$_2$ in DI water was used to etch the Au layer, and a mixture of phosphoric acid and DI water was used to etch the YBCO thin film. Photoconductive switches and coplanar waveguides were fabricated on sapphire substrates with pre-patterned YBCO thin films using the same method as discussed above. A contact resistance of ≈ 2.5 Ω for each contact was obtained.

In both cases, the negligible contact resistance is critical for the measurements shown in this work. In measurements of critical current $J_c$, this is important to avoid heating at contacts, which undermines the value of $J_c$. In measurements of depairing current $J_c^*$, this guarantees that the observed dynamics are from the sample itself instead of the change of contact resistance.

## S3. Characterization of NbN device

An optical microscope image of the NbN ultrafast device used for the measurements is shown in Figure S3. The sample size is, width ≈ 10 um × length ≈ 30 um × thickness ≈ 20 nm.

In order to directly compare the conventional critical current density $J_c$ and the intrinsic depairing current density $J_c^*$, both DC and picosecond transport measurements were conducted on the same device.

In DC transport measurements, there were three contributions to the measured resistance: Au signal line of the coplanar waveguide, contact area and sample. To eliminate the contribution from the Au signal line, during Ti/Au deposition for fabricating the coplanar waveguides, a narrow line was deposited on another MgO substrate. This narrow line was separately characterized with a Physical Property Measurement System (PPMS) to extract its temperature-dependent resistivity. This served as a reference to calculate the resistance contribution of the Au signal line on the ultrafast device (yellow curve in Fig. S4a). The resistance versus temperature dependence is shown in Fig. S4b after subtracting this contribution. At 7 K, below the superconducting transition temperature, the measured resistance is ≈ 6 Ω, with ≈ 3 Ω from the left and right contact. In Fig. S5, the sample resistance $R_{sample}$ is presented after subtracting the contact resistance.

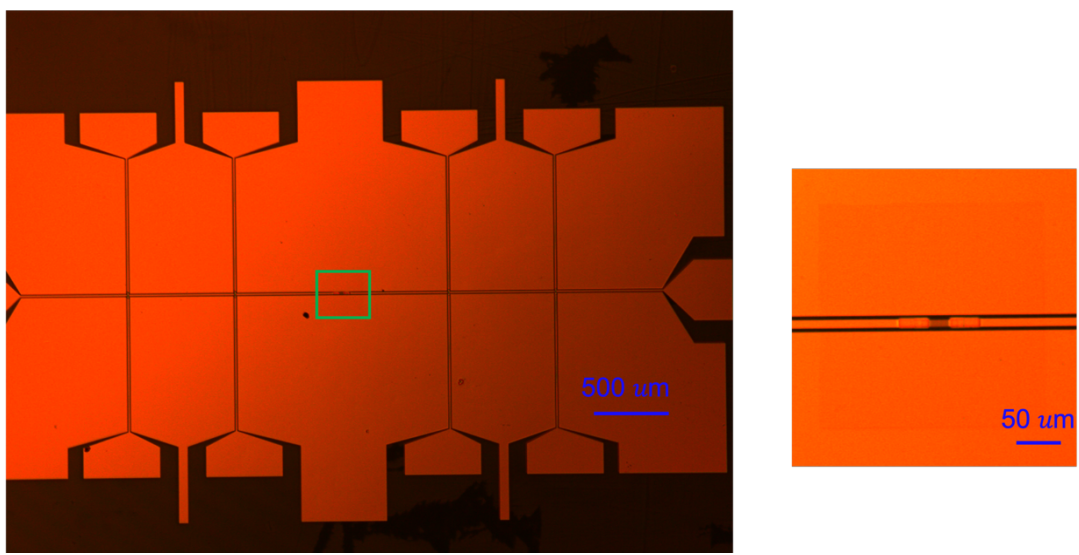

**Figure S3| Optical micrograph of NbN ultrafast device.**

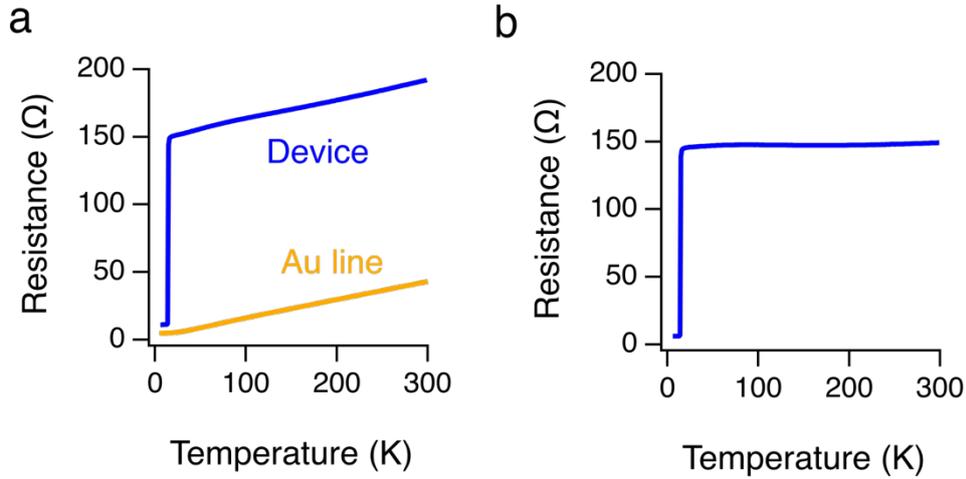

**Figure S4| Resistance contribution from the Au signal line of the coplanar waveguide.** a, Device resistance (blue) and calculated Au line resistance (yellow) versus temperature. b, Resistance versus temperature after subtracting the contribution of the Au line.

To exclude possible heating effects in the DC transport measurement of the conventional critical current density $J_c$, we used the measurement configuration shown in Fig. S6. A Keithley 6221 current source and a Keithley 2182 nanovoltmeter were combined to launch µs-long current pulses. The measurement of the sample resistance was conducted in a fixed 400 µs time window with a tunable delay after the onset of the applied current. As shown in Fig. S4b, the critical current value — at which sample resistance starts rising — remained constant when increasing the delay time. Rather, the sample resistance increased faster after the current was above the critical value, due to the accumulated heating after the sample became dissipative. In this way, we can exclude a possible heating effect at the contacts while measuring the DC critical current density.

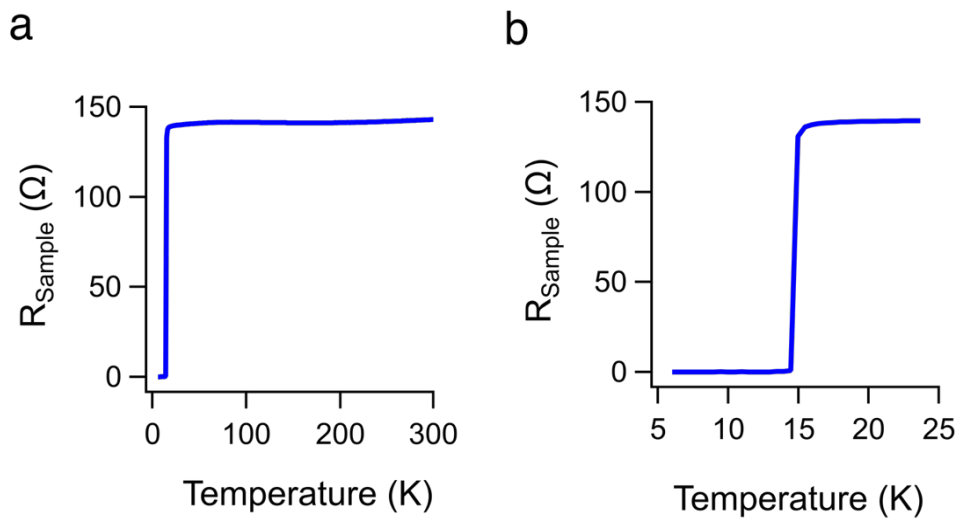

**Figure S5| Superconducting transition of NbN sample.** a, b, Sample resistance versus temperature.

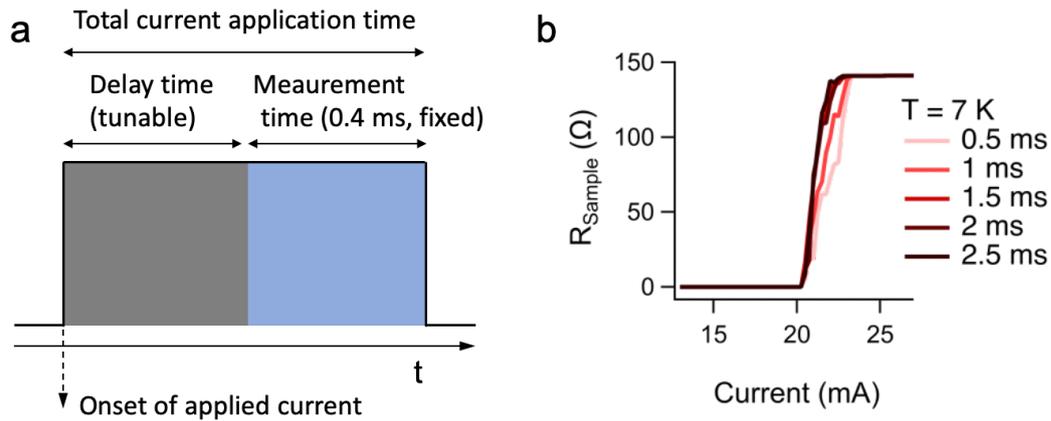

Figure S6| **Dependence of the DC critical current measurements on delay time.** a, Configuration of measurement setup. b, NbN sample resistance versus current for five different delay times.

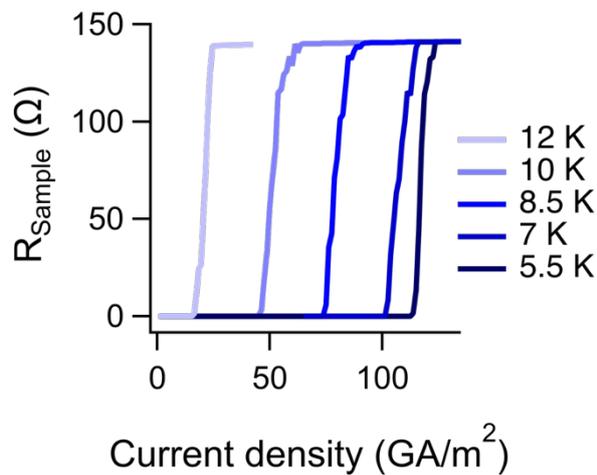

Figure S7| **DC critical current density measurements of the NbN sample at different temperatures.**

## S4. Characterization of YBCO devices

An optical microscope image of the YBCO ultrafast device used for the measurements is shown in Figure S8. The sample size is, width ≈ 10 um × length ≈ 30 um × thickness ≈ 50 nm.

The same procedures as described for the NbN device were performed to identify the resistance contribution from the Au signal line, contact and sample (Figures S9 and S10). The resistance of each contact is ≈ 2.5 Ω. In Fig. 10b the sample resistance $R_{sample}$ is presented after subtracting the contact resistance. Compared with measurements on NbN, the measurement temperature is at 50 K higher than 7 K and the contact resistance is smaller, so we expect there is also no heating effect in the measurement of DC critical current density $J_c$.

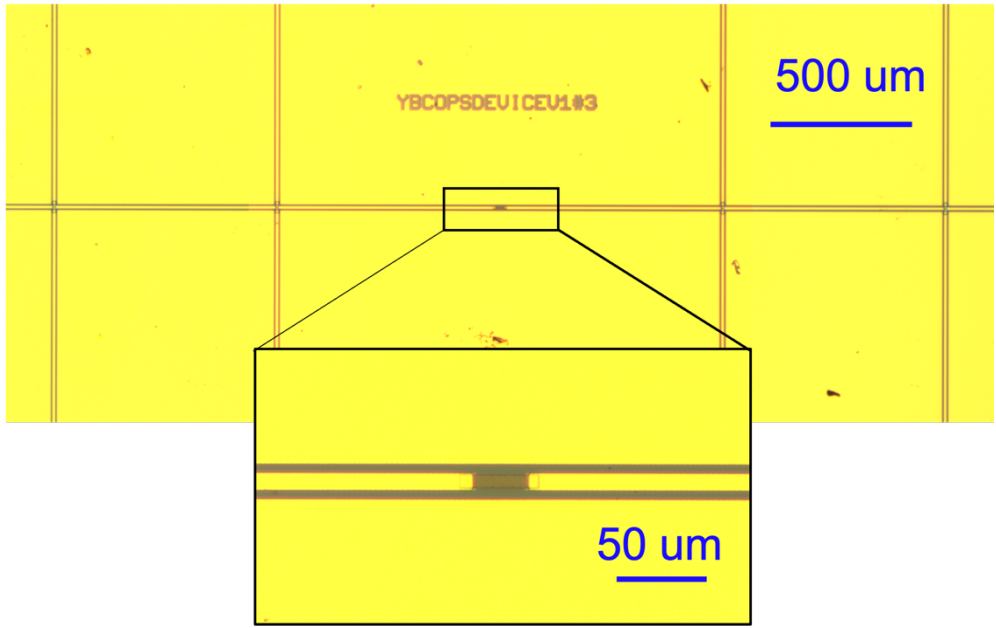

**Figure S8| Optical micrograph of YBCO ultrafast device.**

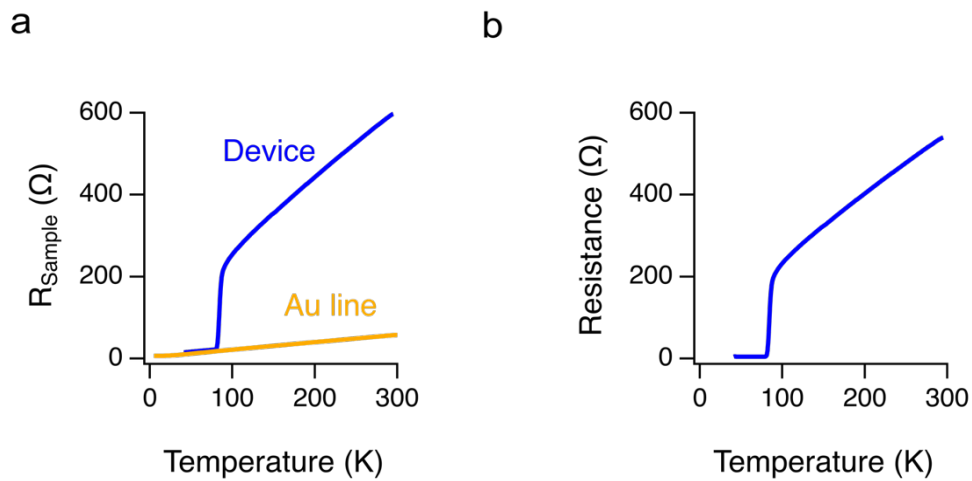

**Figure S9| Resistance contribution from the Au signal line of the coplanar waveguide.** a, Device resistance (blue) and calculated Au line resistance (yellow) versus temperature. b, Calibrated resistance versus temperature after subtracting the contribution of the Au line.

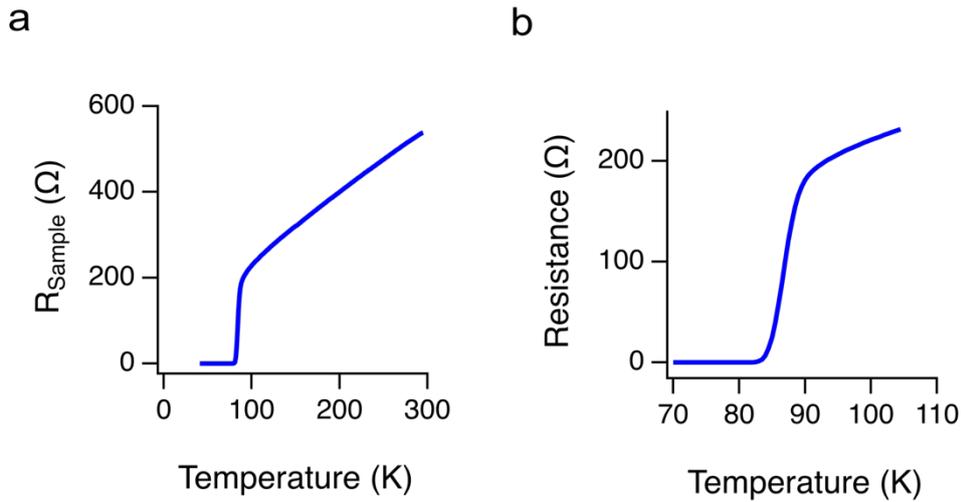

**Figure S10| Superconducting transition of the YBCO sample.** a, b, Sample resistance versus temperature.

## S5. Calibration procedure

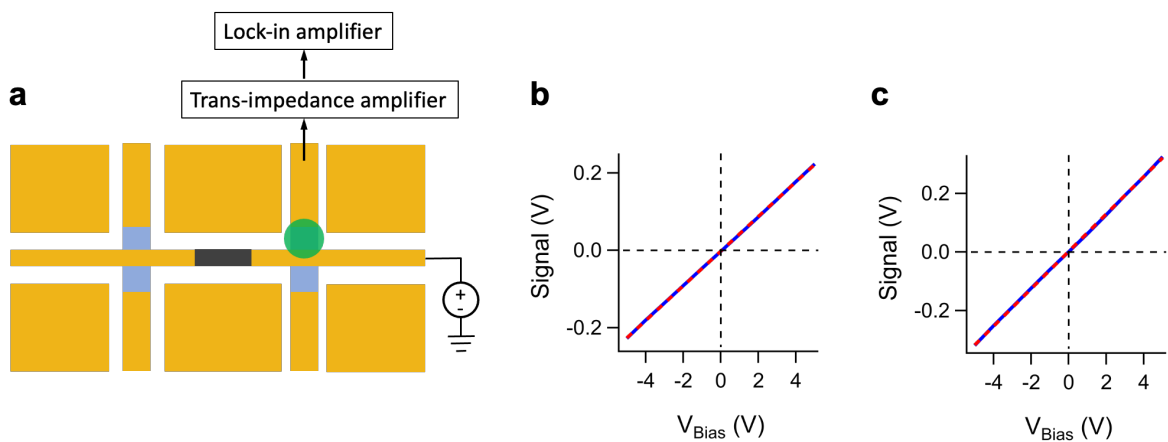

**Figure S11| Switch calibration.** a, Sketch for calibration process. b, Calibration of the photoconductive switch used for sampling incoming/reflective pulses. The measured curve is shown in blue and the linear fitting in dashed red. c, Calibration for the photoconductive switch used for sampling the transmitted pulses. The measured curve is shown in blue and the linear fitting in dashed red.

The calibration procedures are discussed in detail in Ref. [2, 3]. Here the sensitivity of each switch used to probe the current pulses was calibrated with the configuration shown in Fig. S11a. The signal line was biased with a voltage source, leaving the other end open. The switch that was calibrated was illuminated with a 1kHz-chopped 515-nm laser beam and the signal was measured with a lock-in amplifier after a custom-built ultralow-noise trans-impedance amplifier. A linear signal versus $V_{bias}$ dependence was obtained. For example, calibrations performed for the switch measuring the incoming/reflected and the transmitted current pulses at 7 K are shown in Fig. S11b and Fig. S11c, respectively. All measured quantities using photo-excited switches were normalized with the slope α of the calibration curve.

The peak current amplitudes of the incoming and transmitted pulses were calculated as follows,

$$I_{peak} = 1.414 \times \frac{V_{signal,peak}}{\alpha \cdot Z_0 \cdot 0.85} \quad (1)$$

Here, $V_{signal,\,peak}$ is the measured peak signal from the amplifier, α is the slope of the switch calibration curve as discussed above, $Z_0$ is the wave impedance of the coplanar guide ≈ 50 Ω, the factor 0.85 is taking into account the average value between the pulses reaching the photoconductive switches and the pulses transmitting through the switches (≈ 30% reflection due to local impedance mismatch, see Ref. [2], which gives (1+0.7)/2 = 0.85) while the prefactor 1.414 is taking into account the correlation effect between the actual current pulse time profile and the switch response time profile, which are assumed to be very similar due to the high bandwidth of the coplanar waveguide (0.7 THz with -3dB), compared with the full half maximum width of current pulse (FHMW ≈ 2 ps).

We have validated the above calibration method by applying a bias current $I_{bias} < I_c$ and checked how much the depairing process shifted. In Fig. S12 data with $I_{bias}$ = 19.25 mA is shown. As can be seen the shift of the transmitted peak current is around 19.45 mA, which validates our calibration method for the peak currents of picosecond pulses and also indicates the different origins of $J_c$ and $J_c^*$.

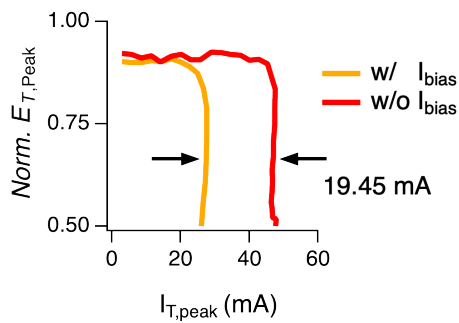

**Figure S12| Offset of the depairing process by applying the bias current $I_{bias}$.**

## S6. Picosecond ultrafast transport measurements on YBCO

The picosecond ultrafast transport measurements on YBCO are shown in Figure S13. The results are similar to the ones on NbN, demonstrating the inductive response at $T$ = 40 K and the resistive response at T = 110 K, below and above $T_c$ = 85 K, respectively. By increasing the incoming peak current density, the YBCO sample also developed a resistive response. However, this transition is a slow and gradual process, as shown in Fig. 4c and 4d. For example, with $J_{T,peak}$ reaching 6 × $J_c$, the normalized transmitted pulse (peak ≈ 43%) is stronger than the one measured at $T$ = 110 K (peak ≈ 23%), right above $T_c$, indicating incomplete suppression of superconductivity.

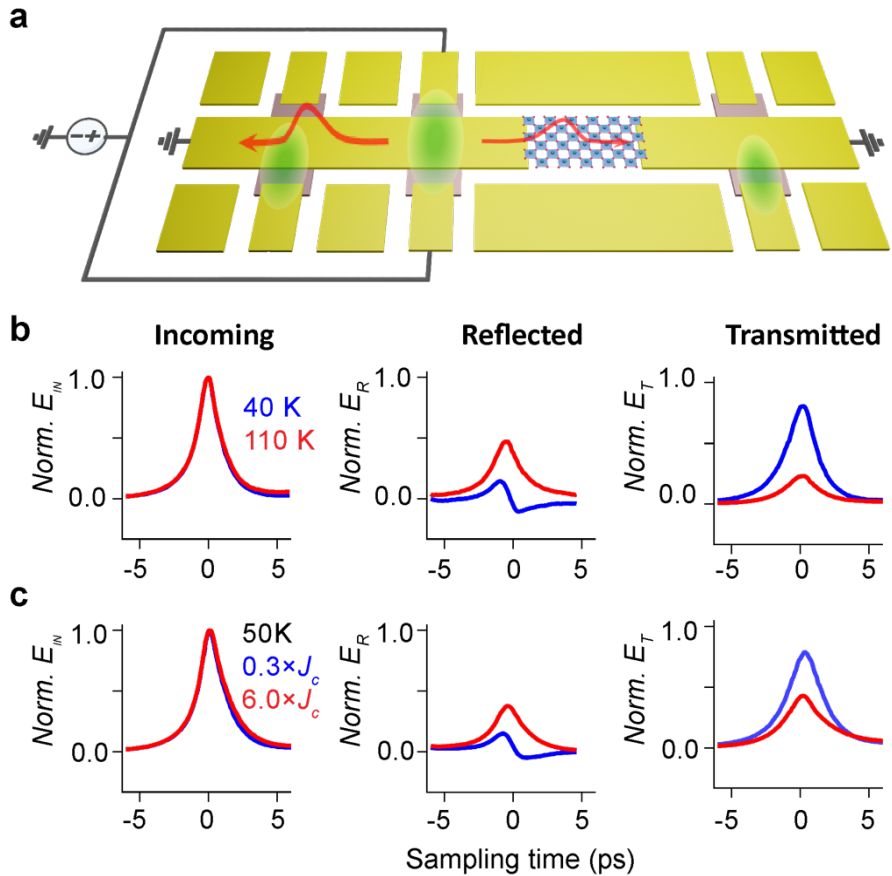

**Figure S13| Picosecond ultrafast electrical transport measurements on YBCO.** a, Illustration of the device architecture, which is the same as the one shown in Fig. 3 of the main text. b, Measurements at 40 K and 110 K, below and above $T_c$ = 85 K, respectively. At 40 K, the reflected pulses show an inductive feature and ~80% of the incoming pulses transmits through. At 110 K, the incoming pulses partially reflect and partially transmit through, due to sample's resistive response. c, Measurements at 50 K with different peak current densities of the incoming pulses. Here the labels indicate the peak current densities of the transmitted pulses for direct comparison. At 6.0 × $J_c$, the sample shows a resistive response, but with a larger transmittance.

## S7. Temperature dependence of $J_c$ and $J_c^*$ in NbN thin film

The temperature dependence of $J_c$ is extracted from Fig. S7. More picosecond transport measurements on NbN at different temperatures are shown in Figure S14. The temperature dependence of $J_c^*$ is extracted from Fig. S14d.

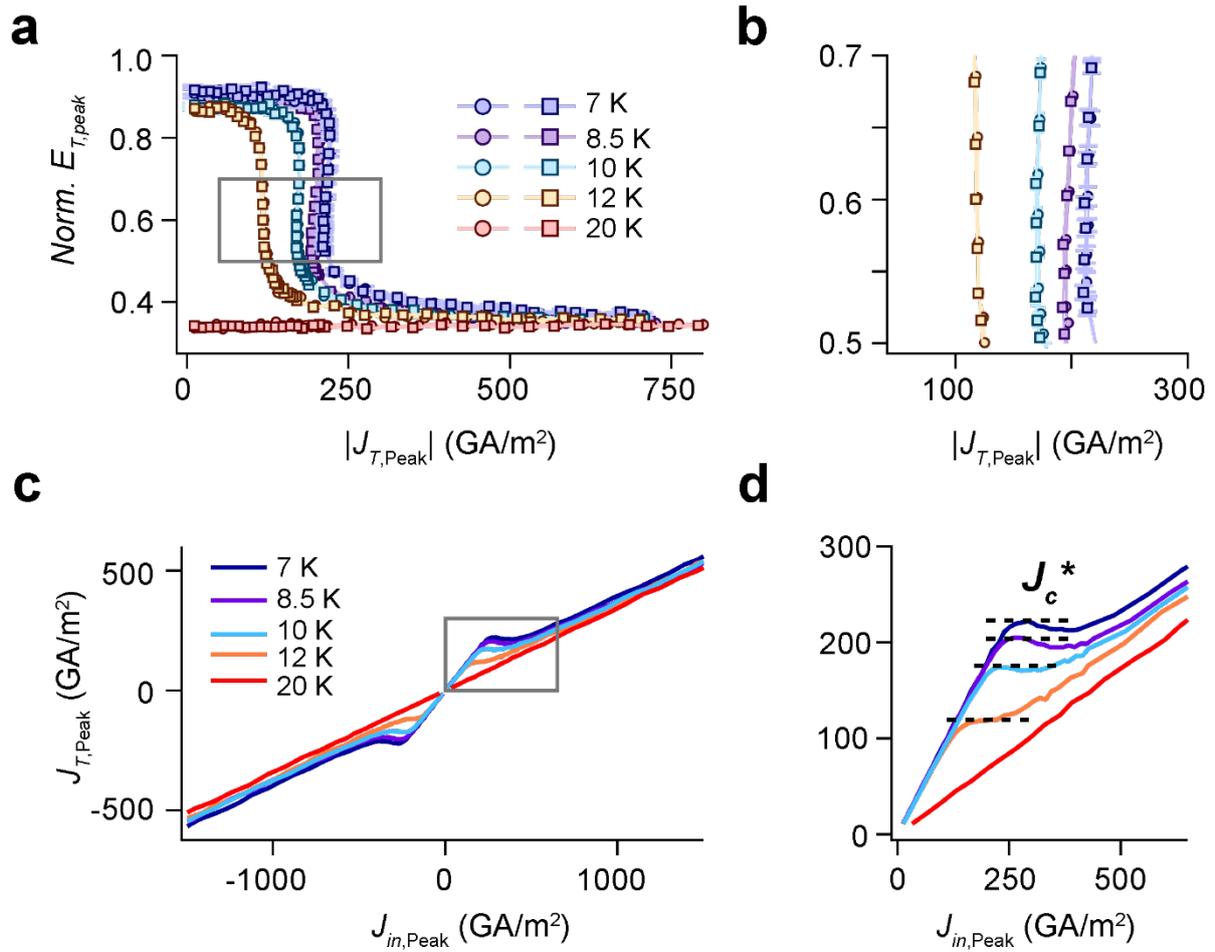

**Figure S14| Temperature dependence of depairing current density $J_c^*$ in NbN.** a, normalized transmitted electric peak field (normalized by the peak electric field of the incoming pulse) versus the transmitted peak current densities at 7, 8.5, 10, 12, 20 K. ● and ■ symbols represent the cases with positive and negative picosecond pulses. b, Zoom-in plot of (a) to show the sharp drop feature. c, transmitted peak current density versus incoming peak current density at 7, 8.5, 10, 12, 20 K. d, Zoom-in plot of (c). Depairing current densities $J_c^*$ at different temperatures are indicated by dashed lines.

## S8. Extraction of superconducting coherence length (ξ), magnetic penetration depth (λ) and Pearl length $P_L$ of NbN strip

### S8.1 Extraction of superconducting coherence length (ξ)

In order to extract the coherence length $\xi(0)$, we measured the superconducting transition in a magnetic field in order to extract $B_{c2}(0)$, from which $\xi(0)$ can be directly calculated. The results are shown in Fig. S15. We took the magnetic field when the sample resistance reaches half of the normal state resistance as $B_{c2}$ and follow the Werthamer-Helfand-Hohenberg (WHH) theory [4].

$$B_{c2}(0) = -0.69 T_c \frac{dB_{c2}}{dT}\Big|_{T=T_c} \qquad (2)$$

$$\xi(0) = \left[\frac{\Phi_0}{2\pi B_{c2}(0)}\right]^{1/2} \qquad (3)$$

We extract $B_{c2}(0) = 12\,T$ and $\xi(0) = 5.24\,nm$, close to $5.3\,nm$ from Ref. [5].

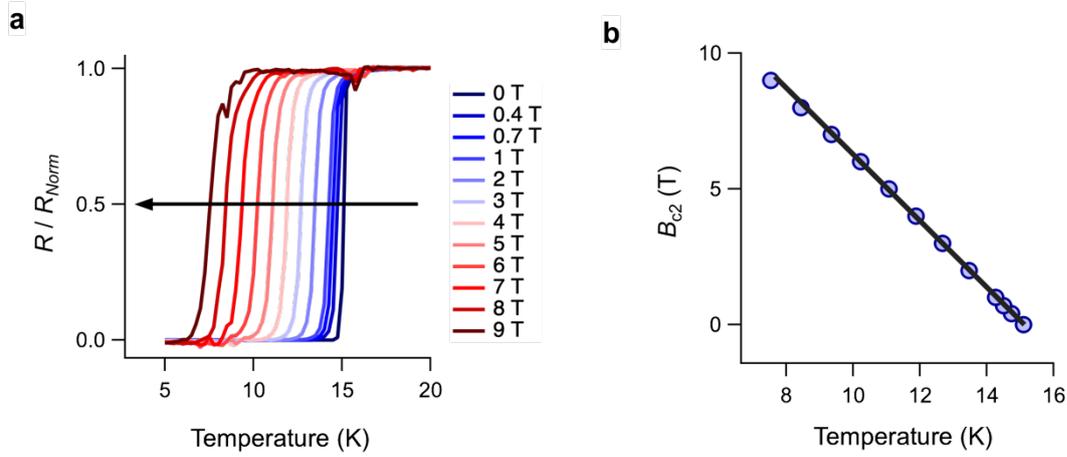

**Figure S15| Measurements of $B_{c2}$ at different temperatures.** (a) Superconducting transition for various applied magnetic fields. (b) Extracted $B_{c2}$ values versus temperature.

### S8.2 Extraction of magnetic penetration depth (λ)

Regarding the magnetic penetration depth $\lambda(0)$, we extracted its value from the sample's kinetic inductance from the equation below [6].

$$L_{kin} = \frac{m}{2n_s e^2} \cdot \frac{l}{s} = \frac{\mu_0 \lambda^2 l}{s} \qquad (4)$$

Where $m$ is the effective mass of the electron; $n_s$ is the superfluid density; $e$ is electron change; $l$ is the length of strip; $s$ is the cross section of the strip; m₀ is the vacuum magnetic permeability; and $\lambda$ is the magnetic penetration depth.

The kinetic inductance of the sample is extracted from the measured reflected pulse shown in Fig. 3b of the main text. Fig. S16 shows the data and the simulated reflected signal with $L_{kin}(0) = 23\,pH$, with matching amplitude. This corresponds to $\lambda(0) = 350\,nm$, which is close to the value of $360\,nm$ in Ref. [5]. The detailed simulation procedure is described below.

The equivalent circuit of the sample together with the coplanar waveguide is shown in Fig. S17. The sample is treated using a two-fluid model. $R_n$ and $L_n$ are the resistance and kinetic inductance of the normal channel; $L_s$ is the kinetic inductance of the superconducting channel and $Z_0$ is the impedance of the coplanar waveguide, which is 50 Ω. $R_n$, $L_n$ and $L_s$ are directly related with the superfluid density as $R_n = R_{n0}/(1-n_s)$, $L_n = L_{n0}/(1-n_s)$ and $L_s = L_{s0}/n_s$, , where $R_{n0}$ and $L_{n0}$ are the sample's resistance and kinetic inductance when all carriers are normal quasiparticles, and $L_{s0}$ is the sample's kinetic inductance when all carrier are Cooper pairs. We take $R_{n0} = 140\ \Omega$, from DC transport measurement. $L_{n0} = L_{s0}$ is treated as a variable.

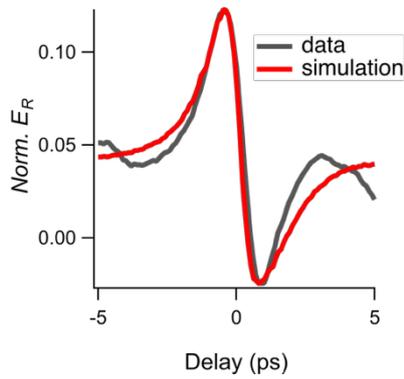

**Figure S16| Simulation of the reflected pulse on NbN at 7 K.**

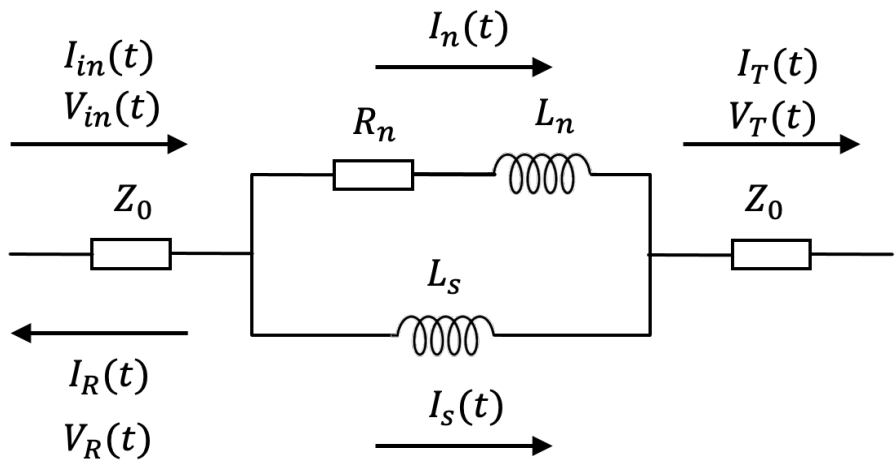

**Figure S17| Circuit model of the sample.**

The reflection and transmission of current pulses at the sample are described by the equations below.

$$I_{in}(t) - I_R(t) = I_s(t) + I_n(t) \tag{5}$$

$$I_T(t) = I_s(t) + I_n(t) \tag{6}$$

$$V_{in}(t) + V_R(t) - V_T(t) = L_s \cdot \frac{dI_s(t)}{dt} \tag{7}$$

$$V_{in}(t) + V_R(t) - V_T(t) = L_n \cdot \frac{dI_n(t)}{dt} + R_n \cdot I_n(t) \tag{8}$$

Where, $V_{in}(t) = I(t) \cdot Z_0$, $V_R(t) = I_R(t) \cdot Z_0$ and $V_T(t) = I_T(t) \cdot Z_0$.

With the incoming current $I_{in}(t)$ shown in Fig. 3b as input, the reflected current $I_R(t)$ is calculated from the differential equations above. The curve with $L_{s0} = 23\ pH$ fits the data best and is shown in Fig. S16.

### S8.3 Extraction of Pearl length ($P_L$)

In order to check whether the current flowing through sample's cross section is uniform, we calculate the Pearl length ($P_L$), which characterize the current penetration into superconductor thin films [7]. $P_L$ is directly related with λ with $P_L = \frac{2\lambda^2}{d}$, where d is the film thickness. From the extracted value of λ above, we get $P_L(0) = 12.2\ um$ and $P_L(7\ K) = 13\ um$, which is larger than the sample's width. Therefore, the current distribution across the sample is uniform. From the same procedure (see Fig. S16), we get $P_L$ (40 K) = 19.4 $um$ and $P_L$ (50 K) = 21 $um$ for YBCO, therefore the current flowing through YBCO at 50 K is also uniform.

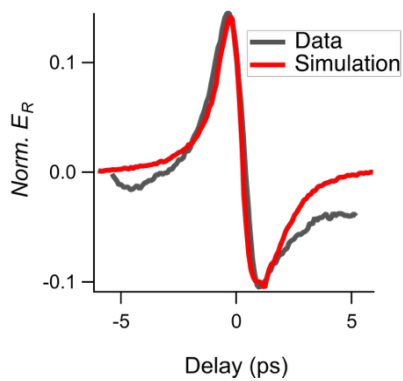

**Figure S18| Simulation of the reflected pulse on YBCO at 40 K.**

## S9. Calculation of $J_c^*$

As the depairing current $J_c^*$ is intrinsically determined by the microscopic properties of the *s*-wave superconductor, it can be estimated from the microscopic superconducting parameters. As shown in the main text, $J_c^* = n_s v_s \cdot (2e)$, where $v_s = \frac{\Delta}{\hbar \cdot k_F} = \frac{\hbar}{\pi m \xi}$. Therefore,

$$J_c^*(T) = \frac{2\hbar e}{\pi m} \frac{n_s(T)}{\xi(T)} \qquad (9)$$

Here, we treat the temperature-dependent superfluid density $n_s(T)$ and coherence length $\xi(T)$ in dirty limit as follows [8]:

$$n_s(T) = \frac{m}{2e^2\mu_0\lambda^2(T)}, \text{ where } \lambda^2(T) = \frac{\lambda(0)^2}{1-(\frac{T}{T_c})^4} \quad (10)$$

$$\xi(T) = \frac{\xi(0)}{\sqrt{\tanh{(1.74\sqrt{\frac{T_c}{T}-1})}}} \quad (11)$$

We take $\lambda(0)$ = 350 nm and $\xi(0)$ = 5.24 nm extracted above. The results are shown in Figure 5 of the main text and exhibit remarkable agreement with the experimental results.

## S10. Theoretical simulations of depairing process

### S10.1 Dynamics model within BCS framework

The dynamics of the superconductor is assumed to be homogeneous throughout the sample geometry. This is confirmed by the Pearl length $P_L = \frac{2\lambda_L^2}{t}$ determined above. In NbN, $P_L \approx 13\ um$ at $T$ = 7 K and in YBCO, $P_L \approx 21\ um$ at $T$ = 50 K, both larger than the width of the sample ~ 10 $um$. Therefore, in both cases, the current distribution is uniform across the sample's cross section.

In discussions below, all quantities are in units with $\hbar = k_B = 1$.

We also assume that the order parameter relaxes sufficiently fast that we can treat the dynamics as essentially adiabatic — that is they follow the instantaneous equilibrium behavior at each point in time as a function of the current. From the Usadel equation [9], we are able to compute the evolution of the supercurrent $J_s$ and gap $\Delta$ as a function of temperature $T/T_c$ and Cooper pair Doppler shift energy $\boldsymbol{Q}_s v_F/T_c$, where $\boldsymbol{Q}_s = \boldsymbol{A} - \frac{\phi_0}{2\pi}\nabla\theta$ is the gauge-invariant momentum of Cooper pairs. The equilibrium Usadel equation is given by

$$[(\epsilon + i\eta)\hat{\tau}_3 - i\Delta\hat{\tau}_2, \hat{g}^R] + ie^2DA^2[\hat{\tau}_3, \hat{g}^R[\hat{\tau}_3, \hat{g}^R]] = 0 \quad (12)$$

where the angled brackets represent a commutator, $\hat{g}^R$ is the spectral function, satisfying the normalization condition $(\hat{g}^R)^2 = 1$, D is the diffusion constant of the material and A is the vector potential, ignoring spatial gradients in the sample geometry. Here $\eta$ represents the Dynes broadening factor, which gives finite quasiparticle life-time due to non-elastic scattering processes (magnetic impurities, phonons, etc.). The gap is determined self-consistently through the gap equation

$$\Delta = -\frac{\pi\lambda}{2} \int_\epsilon tr\, \hat{\tau}^- \hat{g}^K \tag{13}$$

where $\lambda$ is the BCS coupling constant and **tr** represents the trace over Pauli matrices $\hat{\tau}$. Here

$\hat{g}^K = (\hat{g}^R - \hat{g}^A) \tanh\frac{\epsilon}{2T}$ and $\hat{g}^A = -\hat{\tau}_3 (\hat{g}^R)^\dagger \hat{\tau}_3$. Accordingly, we expand the current in the Usadel equation up to first order in time derivatives, capturing the non-linear static response of the superconductor as well as the non-linear static response of the normal state, resulting in a non-linear two-fluid model.

$$J(w) = -n_s(A)A + \sigma(A)E \tag{14}$$

where

$$n_s = i\frac{\pi\sigma_n}{4} \int_\epsilon tr\, [\hat{\tau}_3 \hat{g}^R(\epsilon)\hat{\tau}_3 \hat{g}^K(\epsilon) + \hat{\tau}_3 \hat{g}^A(\epsilon)\hat{\tau}_3 \hat{g}^K(\epsilon)] \tag{15}$$

is the superfluid density and

$$\sigma = -\frac{\pi\sigma_n}{4} \int_\epsilon tr\, \left[\hat{\tau}_3 \hat{g}^R(\epsilon)\hat{\tau}_3 \frac{\partial \hat{g}^K(\epsilon)}{\partial\epsilon} - \hat{\tau}_3 \hat{g}^A(\epsilon)\hat{\tau}_3 \frac{\partial \hat{g}^K(\epsilon)}{\partial\epsilon}\right] \tag{16}$$

is the conductivity. Here $E(t) = -\frac{\partial Q_s}{\partial t}$ is the electric field.

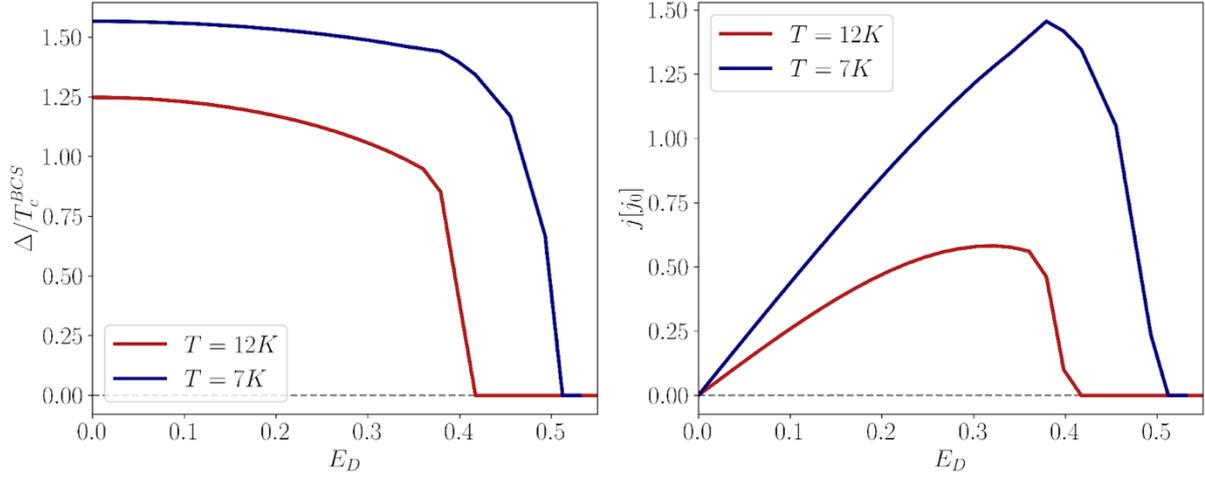

**Figure S19| Nonlinear response of s-wave superconductor under strong current drive.** a, Gap in units of $T_c^{BCS}$ as a function of the Doppler shift $E_D$ for 7 and 12K. b, Normalized supercurrent in units of $j_0$ (see text) as a function of Doppler shift.

### S10.2 Dynamics model within tDGL theory

To simulate the response of the system within tDGL theory, we assume a two fluid model. The superfluid response is given by the usual time-dependent GL equations

$$\tau_{GL}\partial_t f = \left[1 - f^2 - \left(\frac{2\pi\xi}{\phi_0}\right)^2 Q_s^2\right] f + \xi^2 \nabla^2 f \qquad (17)$$

where $f$ is the normalized superconducting gap $f = \Delta(\mathbf{r},t)/\Delta_0$, $\Phi$ is the voltage potential and $\xi$ is the coherence length of the system. Here, $Q_s$ represents the gauge invariant vector potential

$$Q_s = \frac{1}{2e}\nabla\theta - A \qquad (18)$$

and θ is the superfluid phase. The superfluid current and the normal state current within the thin sheet are given as

$$J_s = \frac{d}{\lambda^2}\delta(z)\theta\left(\frac{w}{2} - |y|\right) f^2 Q_s \qquad (19)$$

$$J_n = d\delta(z)\theta\left(\frac{w}{2} - |y|\right)\sigma_n E \qquad (20)$$

and the total current is the summed contributions from superfluid and normal fluid currents.

For the calculations in the tDGL framework, the magnetic penetration lengths are extracted from the measurements $\lambda = 350\ nm$ at 7K and $\lambda = 400\ nm$ at 12K for NbN as discussed in S.8.2. The normal state resistivity is set to match the transmittance at $T > T_c$. We take the microscopically predicted depairing current values at two temperatures as input.

## S10.3 Simulating sample's response under strong ps current drive

The sample's response when incorporated into coplanar waveguide, is modeled under the assumption that the sample is a lumped element, which is valid with sample length ~ 30 $\mu m$ << pulse wavelength ~ 150 $\mu m$. This means that at the sample position, we can write for incoming, reflected and transmitted current/voltage respectively,

$$I_{in}(t) - I_r(t) = I_t(t) = I(t) \tag{21}$$

$$V_{in}(t) + V_r(t) = V_t(t) + V(t) \tag{22}$$

Where $I_{in}(t)$, $I_r(t)$, $I_t(t)$ are the incoming, reflected and transmitted current pulse, $V_{in}(t)$, $V_r(t)$, $V_t(t)$ are the incoming, reflected and transmitted voltage pulse. $I(t)$ and $V(t)$ are the current and voltage across the sample.

The incoming, reflected and transmitted voltage pulses have a linear relation to the current, $V_{in,r,t} = Z_0 I_{in,r,t}$, where $Z_0$ is the wave impedance of coplanar waveguide. Combining everything together we get a relation between the current $I(t)$ and voltage $V(t)$ across the sample:

$$I_{in}(t) = I(t) + \frac{V(t)}{2Z_0} \tag{23}$$

This can be re-cast in the form of a differential equation which can be solve numerically $V(t) = E(t)L = \partial_t Q_s L$, where L is the sample length.

$$\frac{\partial Q_s}{\partial t} = 2Z_0 \left( I_{in}(t) - I(Q_s(t)) \right) \tag{24}$$

The normal state resistivity is set to match the transmittance at $T > T_c$. We choose the BCS coupling constant to match the observed $T_c$ of 14.5 K. Within the dirty superconductor framework and Dynes $\beta$ of zero, the superfluid density is given as

$$\pi \sigma_n \Delta \tanh\left(\frac{\beta \Delta}{2}\right) = n_s \tag{25}$$

Therefore, the samples kinetic inductance within the superconducting phase for low pulse strength is completely determined by the normal state resistivity and gap size

$$L_s = \frac{R_n}{\Delta \tanh\left(\frac{\beta \Delta}{2}\right)} \tag{26}$$

The successful application of our theory at weak pulses proves the validity of the dirty limit approximation and the lumped circuit element model. The only remaining parameter within our framework is then the depairing current. To extract this, we use the experimentally measured value of coherence length at 0K and use the dirty superconductor relation

$$\xi(T) = \sqrt{\frac{D}{2\Delta(T)}} \rightarrow \xi(T) = \xi(0)\sqrt{\frac{\Delta(0)}{\Delta(T)}} \tag{27}$$

Where we obtain the gap as a function of T from the self-consistency equation from above. To relate the coherence length to the depairing current, we go back to the current - vector potential relation and the Usadel equation. We define a dimensionless Doppler shift as $E_D^2 = e^2 D A^2/T_c$ and the full current in the DC limit can be re-written as

$$j = \sigma_n T_c \frac{1}{e\sqrt{D/T_c}} \left[\frac{\pi E_D}{4} \int_\epsilon \text{tr}\left(\hat{\tau}_3 \hat{g}^R(\epsilon, E_D)\hat{\tau}_3 \hat{g}^K(\epsilon, E_D) + \hat{\tau}_3 \hat{g}^A(\epsilon, E_D)\hat{\tau}_3 \hat{g}^K(\epsilon, E_D)\right)\right] \tag{28}$$

Importantly, we see from the Usadel equation that the spectral functions only depend on $E_D$ and do not contain additional factors or dependence on the diffusion constant. Therefore, the depairing current can be re-written as

$$I_c^* = \frac{L}{R_n} \frac{T_c}{\xi(T)} \sqrt{\frac{T_c}{2\Delta(T)}} \times C_0 \tag{29}$$

where $C_0$ is the maximum of the expression in the square bracket above as a function of $E_D$, at fixed temperature. We define $I_0$ as the $I_c^* = I_0 * C_0$. This allows us to extract the critical current from the zero temperature measurements of $B_{c2}$, leaving only one undetermined parameter in our theory which is the Dynes broadening $\eta$. In our simulations, we choose the value for Dynes broadening equal to 8% of the $T_c^{BCS}$, to ensure numerical stability and convergence. The finite value of $\eta$ causes corrections in the gap temperature dependence and critical temperature. Therefore, in our fitting procedure, we choose the BCS coupling to correspond to $T_c(\eta = 0) = T_c^{BCS} = 1$ and represent all the quantities in units of $T_c^{BCS}$. For the chosen value of Dynes broadening we obtain $T_c = 0.89\, T_c^{BCS}$, therefore $T_c^{BCS} = \frac{1}{0.89} * 14.5\, K$ and in the above equation for depairing current we replace $T_c$ by $T_c^{BCS}$, since that is the natural unit of gap.

Below we show the simulations based on microscopic BCS theory in dirty limit and tDGL theory (Fig. S20). As we can see the tDGL cannot capture the sharp drop of transmittance at T = 7 K, while the microscopic theory can reproduce the observed current density threshold ($J_c$*) and the abrupt collapse of superconductivity above this threshold much better. This is because, the Ginzburg-Landau theory relies on the expansion in the power of order parameter and is valid only near $T_c$ [10, 11]. As expected, at $T$ = 12 K, closer to $T_c$, the results from both theories start to coincide.

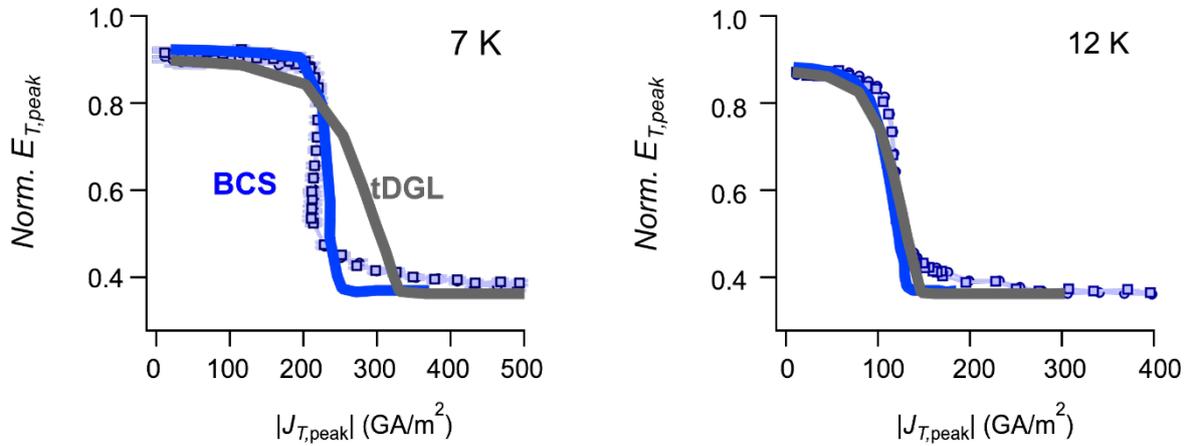

**Figure S20| Comparison of simulation results based on BCS theory in dirty limit and tDGL theory.** The simulation results based on BCS theory in dirty limit and tDGL theory are plotted in blue and grey colors. The measurement data is plotted in light blue.

Our instantaneous theory provides estimates of $J_c^*$ in equilibrium. In a dynamical setting, the value of the maximum current that can be passed through the superfluid might be different. Experimentally, we see that in the case of NbN the dynamical depairing current is very close to the theoretically calculated depairing current as demonstrated in Fig. S20. Since the two values are consistent, we can justify the use of instantaneous equilibrium theory, in which depairing dynamics is faster than the timescale of picosecond current pulse.